\documentclass[a4paper,12pt,dvipdfmx]{article}
\usepackage{amssymb,amsmath}
\usepackage{bm}%
\usepackage{stmaryrd}
\usepackage{braket}
\usepackage{mathtools}
\usepackage{cite}

\usepackage[dvipdfmx]{hyperref}
\hypersetup{
    colorlinks=true,
    citecolor=blue,
    linkcolor=blue,
    urlcolor=orange,
}

\def\nn{\nonumber}
\def\Z2{\mathbb{Z}_2^2}
\def\g{\mathfrak{g}}

\def\hf{\frac{1}{2}}
\def\Min{\mathcal{M}^{(1,1)}_{1}}

\title{Comments of $\mathbb{Z}_2^2$-supersymmetry in superfield formalism}
\author{S. Doi and N. Aizawa %\footnote{Corresponding author: aizawa@p.s.osakafu-u.ac.jp}
	\\[10pt]
Department of Physical Science, Osaka Prefecture University, \\
Nakamozu Campus, Sakai, Osaka 599-8531, Japan}

\date{}

\begin{document}
	\maketitle
	\thispagestyle{empty}
	
	\vfill
\begin{abstract}
 We investigate superfield formulation of the minimal $\mathbb{Z}_2^2$-supersymmetry. 
It is shown that the integrability on $\mathbb{Z}_2^2$-superspace guarantees the invariance of action. Then we present two superfields which carry distinct irreducible representations of the $\mathbb{Z}_2^2$-supersymmetry algebra. 
One of them gives integrable Lagrangian and the other does not. 
We also show that integrable superfields with different $\mathbb{Z}_2^2$-degree also carry irreducible representations and they give invariant actions. 
To perform this analysis, the representation theory of the minimal $\mathbb{Z}_2^2$-supersymmetry algebra is studied in some detail. 
\end{abstract}

\clearpage	
\setcounter{page}{1}
%%%%%%%%%%%%%%%%%%%%%%%%%%%%%%%%%%%%%%%%%%%%%%%%%%%%%%%%%%%
%
%
%
%%%%%%%%%%%%%%%%%%%%%%%%%%%%%%%%%%%%%%%%%%%%%%%%%%%%%%%%%%%

\section{Introduction}

Symmetries generated by higher graded Lie algebras attract renewed interest \cite{tol2,tol,AKTT1,AKTT2,Bruce,BruDup,AAD,AAD2,AKTcl,AKTqu,DoiAi,brusigma,bruSG,StvdJ,ISvdJ,Topp,Topp2,KuTo,Que}. 
One of the reasons for this is that such symmetries were found in simple and fundamental physical systems such as the mixture of parabosons and parafermions \cite{tol}, the non-relativistic version of Dirac equation \cite{AKTT1,AKTT2}, quantum mechanics with multi-component wavefunctions \cite{BruDup} and so on. 
It is surprising that the higher graded symmetries, which have been thought unusual in physics, are recognized in rather simple systems. 
It is also shown that quantum mechanical particles obeying $\Z2$-graded symmetries\footnote{$\Z2 := \mathbb{Z}_2 \times \mathbb{Z}_2.$ } exhibit exotic statistics, namely, they are not bosons nor fermions \cite{Topp,Topp2}. 
This allows us to detect experimentally deviations from the conventional spin-statistics relation.  
Furthermore, a new class of integrable systems is constructed based on $\Z2$-graded supersymmetry ($\Z2$-SUSY for short) \cite{bruSG}. 
Since the $\Z2$-SUSY contains the standard one as a subset, further studies of $\Z2$-SUSY models would provide a candidate of physics beyond the standard model and novel theories in low-dimensional condensed matter physics. 

Superfields are a powerful tool to analyse physics and mathematics of supersymmetry. 
It is therefore natural to extend them to $\Z2$-graded setting \cite{brusigma,KuTo}. 
However, there are some issues on the $\Z2$-superfields. 
Among others we address the following three: 
(1) A $\Z2$-superfield has infinitely many component fields, even for the minimal $\Z2$-SUSY so that it carries an infinite dimensional representation of the $\Z2$-SUSY algebra introduced in \cite{Bruce}. 
To construct models of $\Z2$-SUSY mechanics we have to extract finite dimensional irreducible representations (irreps) form the superfield. 
(2) Irreps of $\Z2$-SUSY  algebras have not been studied systematically. 
We do not know what kind of irreps exist even for the minimal one. 
This makes difficult to consider the extraction of irreps from the superfields. 
(3) $\Z2$-SUSY invariant actions will be obtained by integrating a function of the superfield and its derivatives on the $\Z2$-superspace. 
The integration on $\Z2$-superspace is a highly non-trivial and it requires the integrability condition on functions  \cite{Pz2nint,brusigma}.   
However, it has not been studied the compatibility between the integrability condition and the $\Z2$-SUSY transformation. 
We remark that integration on general $\mathbb{Z}_2^n$-manifold is an open problem and only the case of $\Z2$-manifold is understood \cite{Pz2nint}. 

In the present work, we investigate the addressed three problems in detail for the minimal $\Z2$-SUSY case, i.e., we consider the $\Z2$-SUSY generated by two supercharges $ Q_{10}, Q_{01}$ and the $ \Z2$-SUSY algebra is of four dimension. 
We also give some invariant actions vis superfield when the superfield formulation works well. 

The plan of this paper is as follows. 
In the next section, we review briefly the necessary notions such as $\Z2$-superspace, functions on it, derivatives and integrals. 
\S \ref{SEC:superfields} is also a review of $\Z2$-SUSY transformation and $\Z2$-superfields. 
In \S \ref{SEC:InvInt}, compatibility of the integrability condition and $\Z2$-SUSY transformation is studied and it is shown that the integrability guarantees the $\Z2$-SUSY invariance of the action. 
We consider a superfield presentation of irreps of the minimal $\Z2$-SUSY algebra in \S \ref{SEC:IrrepInteg}. 
As proved in Appendix A, the $\Z2$-SUSY algebra has two types of four dimensional irreps. 
We present two different constraints on the superfield which reproduce the two types of irreps. 
One constraint is identical to the $z$-constraint by Bruce \cite{brusigma} and one may obtain $\Z2$-SUSY invariant action via the constrained superfield. 
While the other novel one reproduces the irrep correctly, but not suitable to construct invariant actions since functions of the superfield are incompatible with the integrability. 
Therefore, we see that the superfield formulation of $\Z2$-SUSY invariant theory is not so trivial problem. 

In Appendix B, it is shown that the irreps carried by the $z$-constrained superfields with non-trivial $\Z2$-degree are obtained from the one in \S  \ref{SEC:Irreps} by the dressing transformation. The $ \Z2$-SUSY invariant actions obtained via the  superfields are also presented. 

%%%%%%%%%%%%%%%%%%%%%%%%%%%%%%%%%%%%%%%%%%%%%%%%%%%%%%%%%%%
%
%
%
%%%%%%%%%%%%%%%%%%%%%%%%%%%%%%%%%%%%%%%%%%%%%%%%%%%%%%%%%%%
\section{Derivatives and integrals on $\Z2$-superspace} \label{SEC:DerInt}

In this section, we present a brief review on functions on a $\Z2$-superspace, their derivatives and integration.

Let us first recall the definition of $\Z2$-graded Lie algebras \cite{RW1,RW2} (see also \cite{Ree,sch1}). 
A $\Z2$-graded vector space is a vector space (over $\mathbb{R}$ or $ \mathbb{C}$) that is the direct sum of homogeneous vector spaces 
\[
  \g = \g_{(0,0)} \oplus \g_{(1,1)} \oplus \g_{(1,0)} \oplus \g_{(0,1)}.
\]
An element of $ \g_{\vec{a}}$  is said to have the $\Z2$-degree $ \vec{a}  \in \Z2.$ 
We define the $\Z2$-Lie bracket by
\begin{equation}
   \llbracket X, Y \rrbracket = XY - (-1)^{\vec{a}\cdot\vec{b}} Y X,
   \quad
   X \in \g_{\vec{a}}, \ Y \in \g_{\vec{b}}
\end{equation}
where $ \vec{a}\cdot\vec{b} $ is the standard scalar product of two dimensional vectors. 
Namely, the $\Z2$-Lie bracket is the commutator (anti-commutator) for $ \vec{a}\cdot\vec{b} $ is even (odd). 
A $\Z2$-graded vector space is said to be a $\Z2$-graded Lie algebra if 
$ \llbracket X, Y \rrbracket \in \g_{\vec{a}+\vec{b}} $ and the Jacobi identity is satisfied:
\[
  \llbracket X, \llbracket Y, Z \rrbracket \rrbracket
  = \llbracket \llbracket X, Y \rrbracket, Z \rrbracket
  + (-1)^{\vec{a}\cdot \vec{b}} \llbracket Y, \llbracket X, Z \rrbracket \rrbracket.
\]
If $ \llbracket X, Y \rrbracket = 0, $ we say that $ X $ and $Y$ are $\Z2$-commutative. 
We also define the even and odd subspaces of $\g$ by $ \g_{(0,0)} \oplus \g_{(1,1)} $ and $\g_{(1,0)} \oplus \g_{(0,1)},$ respectively.  

 The $\Z2$-superspace was introduced by Bruce as  an extension of superspace \cite{Bruce}. It is a coordinate space spanned by variables with $\Z2$-degree and all the variables are $\Z2$-commutative. In contrast to the ordinary superspace, the relationship between two odd variables are not necessarily anti-commutative as $ [\g_{(1,0)}, \g_{(0,1)}] = 0.$   
We now consider a $\Z2$-superspace with the following coordinate system
\begin{equation}
 \mathcal{M}^{(1,1)}_{1} := \{ \underbrace{t}_{(0,0)},\underbrace{z}_{(1,1)},\underbrace{\theta_{10}}_{(1,0)},\underbrace{\theta_{01}}_{(0,1)}\}.
\end{equation}
We call this a $d=1,\mathcal{N}=(1,1)\quad \Z2$-Minkowski space. 
Note that $ \theta_{10}^2 = \theta_{01}^2 = 0. $ 

Now we consider a function $ F(t,z,\theta_{10},\theta_{01}) $ on $\mathcal{M}^{(1,1)}_{1}$. In general, a function on a $\Z2$-superspace has $\Z2$-degree, but we do not need to specify the degree, since the discussion in this section is independent of it. 
One may expand $F(t,z,\theta_{10},\theta_{01})$ in a (formal) power series of $ \theta_{10}, \theta_{01}$ and $z$
\begin{equation} \label{Fcomp}
F(t,z,\theta_{10},\theta_{01})=\sum_{k=0}^{\infty}\sum_{\alpha,\beta=0,1} z^{k}\,\theta_{10}^{\alpha}\,\theta_{01}^{\beta}\, F_{k\alpha\beta}(t).
\end{equation}
The derivative of $ F(t,z,\theta_{10},\theta_{01}) $ with respect to a variable with non-trivial $\Z2$-degree follows immediately from the definition
\begin{align}
\frac{\partial}{\partial \theta_{10}} \theta_{10} &= \frac{\partial}{\partial \theta_{01}} \theta_{01}=1, \nn \\
\frac{\partial}{\partial z}z^{k}&=kz^{k-1},\qquad \text{for all} \quad k>0 .
\end{align}
We remark that the differential operators also have the $\Z2$-degree.

On the other hand, integration on $ \mathcal{M}^{(1,1)}_{1} $ is highly non-trivial \cite{Pz2nint} (see also \cite{brusigma}). 
According to \cite{Pz2nint} we define a Berezinian section by
\begin{equation}
  \sigma := dt\, dz\, d\theta_{10}\,  d\theta_{01} F(t,z,\theta_{10},\theta_{01}) 
\end{equation}
and $\sigma$ is said to be \textit{integrable} if its components do not contain the monomial $ z F_{100}(t). $ 
Then the integral of $\sigma $ on $ \mathcal{M}^{(1,1)}_{1} $ is defined by
\begin{align} \label{IntDef}
\int_{\mathcal{M}^{(1,1)}_{1}} \sigma &= \int_{\mathcal{M}^{(1,1)}_{1}} dt\, dz\, \partial_{\theta_{10}}\, \partial_{\theta_{01}} 
  \sum_{k=0}^{\infty}\sum_{\alpha,\beta=0,1}z^{k}\,\theta_{10}^{\alpha}\,\theta_{01}^{\beta}\, F_{k\alpha\beta}(t) \nn \\
&=\int_{\mathbb{R}} dt\, F_{011}(t).
\end{align}
Roughly speaking, the integrals with respect to the odd coordinates are identical to the standard Grassmann integral and that with respect to $z$ is defined by
\begin{equation}
  \int dz\, z^{k} = \delta_{k,0}.
\end{equation} 
The integrability condition on $\sigma $ is necessary to make the integral \eqref{IntDef} well-defined, that is, any coordinate changes in $ \Min $ preserve the value of the integral.

 %%%%%%%%%%%%%%%%%%%%%%%%%%%%%%%%%%%%%%%%%%%%%%%%%%%%%%%%%%%%%%%%%%%%%%
\section{Superfields and invariance of action}
\subsection{Superfields} \label{SEC:superfields}
\setcounter{equation}{0}

The $\Z2$-supertranslation acting on $\Min$ is defined by \cite{Bruce}
\begin{align}
t'&=t+a+i(\epsilon_{10} \theta_{10}+\epsilon_{01}\theta_{01}), &
z'&=z+\mu- (\epsilon_{10}\theta_{01}- \epsilon_{01}\theta_{10}), \nn \\
\theta'_{10}&= \theta_{10} + \epsilon_{10}, &
\theta'_{01} &= \theta_{01} + \epsilon_{01}. \label{dfnZ22supertrans}
\end{align}
where $a,\mu,\epsilon_{10},\epsilon_{01}$ are parameters with $\Z2$-degree
\begin{equation}
\{ \underbrace{a}_{(0,0)},\underbrace{\mu}_{(1,1)},\underbrace{\epsilon_{10}}_{(1,0)},\underbrace{\epsilon_{01}}_{(0,1)}\}.
\end{equation}
These infinitesimal translations are generated by the following operators
\begin{align}
H&=i\frac{\partial}{\partial t}, & Z&=i\frac{\partial}{\partial z}, \nn \\
Q_{10}&= i\theta_{10} \frac{\partial}{\partial t}+\frac{\partial}{\partial \theta_{10}}-\theta_{01} \frac{\partial}{\partial z}, &
Q_{01}&= i\theta_{01} \frac{\partial}{\partial t}+\frac{\partial}{\partial \theta_{01}}+\theta_{10} \frac{\partial}{\partial z}. \label{Z22SUSYalg}
\end{align}
These operators close in the $\Z2$-SUSY algebra ($d=1 \ \Z2$-super Poincar\'e algebra) 
\begin{align}
\{Q_{10},Q_{10}\}&=\{Q_{01},Q_{01}\}=2H ,& [H,Q_{10}]&=[H,Q_{01}]=0, \nonumber \\
[Q_{01},Q_{10}]&=2iZ ,& [H,Z]&=\{Q_{10},Z\}=\{Q_{01},Z\}=0. \label{defz22spa}
\end{align}

If a function on $\mathcal{M}^{(1,1)}_{1}$ is a scalar under the $\Z2$-graded supertranslation $(\ref{dfnZ22supertrans})$, it is called the superfield
\begin{equation}
\Psi' (t',z',\theta'_{10},\theta'_{01})=\Psi (t,z,\theta_{10},\theta_{01}).
\end{equation}
Contrary to the standard superfields, the $\Z2$ version of superfield have an infinite number of component fields
\begin{align} \label{CompExpansion}
\Psi (t,z,\theta_{10},\theta_{01}) = \sum_{k=0}^{\infty}\sum_{\alpha,\beta=0,1}  z^{k}\theta_{10}^{\alpha} \theta_{01}^{\beta}f_{k\alpha\beta}(t). 
\end{align}
This means that $ \Psi $ carries an infinite dimensional representation of the $\Z2$-SUSY algebra \eqref{defz22spa}.  
We thus need to extract a finite dimensional irreducible representation  which is discussed in the next section. 
Furthermore, it may have a non-trivial $\Z2$-degree which we denote $ (\Delta_{1},\Delta_{2})$. 
The $\Z2$-degree of the component field $ f_{k\alpha\beta}(t) $ is given by 
$( \Delta_1+ k+\alpha, \Delta_2+k+\beta). $ 

The $\Z2$-SUSY transformation of the component fields is obtained in the same way as the standard SUSY theories
\begin{align} %
\delta \Psi =&\Psi' (t,z,\theta_{10},\theta_{01})-\Psi (t,z,\theta_{10},\theta_{01}) \nn \\
=& -(\epsilon_{10}Q_{10}+\epsilon_{01}Q_{01})\Psi
\nn \\
=&\sum_{k=0}^{\infty} z^{k}\big\{ (-1)^{k+\Delta_{1}}f_{k10}\epsilon_{10}+(-1)^{k+\Delta_{2}}f_{k01}\epsilon_{01} \big\} \nn \\
&+\sum_{k=0}^{\infty}  z^{k} \theta_{10}\big[ (-1)^{k+\Delta_{1}}i \dot{f}_{k00}\epsilon_{10}+(-1)^{k+\Delta_{2}} \{ f_{k11}+ (k+1) f_{k+100}\}\epsilon_{01} \big]  \nonumber \\
&+\sum_{k=0}^{\infty} z^{k}\theta_{01}\big[ (-1)^{k+\Delta_{1}}\{ f_{k11}- (k+1) f_{k+100}\}\epsilon_{10}+(-1)^{k+\Delta_{2}}i \dot{f}_{k00}\epsilon_{01} \big]  \nonumber \\
&+ \sum_{k=0}^{\infty}z^{k}\theta_{10}\theta_{01}\big[ (-1)^{k+\Delta_{1}}\{ i\dot{f}_{k01}- (k+1) f_{k+110}\}\epsilon_{10}
 \nn \\
 & \hspace{50mm }
  + (-1)^{k+\Delta_{2}}\{ i\dot{f}_{k10}+ (k+1) f_{k+101}\}\epsilon_{01} \big]
 \nn \\
 \equiv & \sum_{k=0}^{\infty}\sum_{\alpha,\beta=0,1} z^{k}\,\theta_{10}^{\alpha}\,\theta_{01}^{\beta}\, \delta f_{k\alpha\beta}
\end{align}
where $ \dot{f} := \frac{df}{dt}. $  
The transformation of component fields is given explicitly by
\begin{align}
\delta f_{k00} &=  (-1)^{k+\Delta_{1}}f_{k10}\epsilon_{10}+(-1)^{k+\Delta_{2}}f_{k01}\epsilon_{01}  \nonumber \\ 
\delta f_{k10} &=  (-1)^{k+\Delta_{1}}i \dot{f}_{k00}\epsilon_{10}+ (-1)^{k+\Delta_{2}}\{ f_{k11}+ (k+1) f_{k+100}\}\epsilon_{01}   \nonumber \\
\delta f_{k01} &= (-1)^{k+\Delta_{1}} \{ f_{k11}- (k+1) f_{k+100}\}\epsilon_{10}+(-1)^{k+\Delta_{2}}i \dot{f}_{k00}\epsilon_{01}  \nonumber \\
\delta f_{k11} &= (-1)^{k+\Delta_{1}}\{ i\dot{f}_{k01}- (k+1) f_{k+110}\}\epsilon_{10}
+(-1)^{k+\Delta_{2}}  \{ i\dot{f}_{k10}+ (k+1) f_{k+101}\}\epsilon_{01} \label{CompFieldsTrans}.
\end{align}
These equations contain $\Delta_{1}, \Delta_{2}$ so that the transformation of the component fields depends on the $\Z2$-degree of superfield.

 %%%%%%%%%%%%%%%%%%%%%%%%%%%%%%%%%%%%%%%%%%%%%%%%%%%%%%%%%%%%%%%%%%%%%%
\subsection{Invariance of action and integrability} \label{SEC:InvInt}

In the superfield formalism of the standard SUSY theories, the partial derivatives of the odd coordinates of superspace are not SUSY covariant. 
The situation is same for the $\Z2$-graded superfields, we thus introduce the covariant derivative 
\begin{equation}
D_{10} = \frac{\partial}{\partial \theta_{10}}-i\theta_{10}\frac{\partial}{\partial t}+\theta_{01} \frac{\partial}{\partial z} ,\qquad
D_{01} = \frac{\partial}{\partial \theta_{01}}-i\theta_{01}\frac{\partial}{\partial t}-\theta_{10} \frac{\partial}{\partial z}.
\end{equation}
The covariant derivatives, together with the supercharges, satisfy the following relations
\begin{alignat}{2}
\{D_{10},Q_{10}\}&=\{D_{01},Q_{01}\}=0 , &\qquad [D_{10},Q_{01}] &=[D_{01},Q_{10}]=0, 
\nonumber \\
\{D_{10},D_{10}\}&=\{D_{01},D_{01}\}=-2H, & 
[D_{01},D_{10}]&=-2iZ .
\end{alignat}
It then follows that a function $ F(\Psi,D_{10}\Psi,D_{01}\Psi,\cdots) $ with a homogeneous $\Z2$-degree is covariant under the $\Z2$-SUSY transformation \eqref{CompFieldsTrans}. 
More precisely, the function can be expanded in a series of the superspace coordinate
\begin{equation}
F(\Psi,D_{10}\Psi,D_{01}\Psi,\cdots) =F(t,z,\theta_{10},\theta_{01})=\sum_{k=0}^{\infty}\sum_{\alpha,\beta=0,1} z^{k}\,\theta_{10}^{\alpha}\,\theta_{01}^{\beta}\,  F_{k\alpha\beta}(t)
\end{equation}
and the $\Z2$-SUSY transformation of the component field $ F_{k\alpha\beta}(t) $ is also given by \eqref{CompFieldsTrans}.

We want a $\Z2$-supersymmetric action by integrating the function on $\Min.$ 
However, the function is not \textit{a priori} integrable. 
We need to impose the integrability condition $ F_{100}(t) = 0 $ and it is not obvious that the integrability condition and the $\Z2$-SUSY transformation are compatible. 
We now show the compatibility, that is, if the function $F(\Psi,D_{10}\Psi,D_{01}\Psi,\cdots)$ is integrable, then 
\begin{align}
S&=\int dt\, dz\, d\theta_{10}\, d\theta_{01} F(\Psi,D_{10}\Psi,D_{01}\Psi,\cdots) \nn \\
&= \int dt F_{011}(t)
\end{align}
is $ \Z2$-supersymmetric. 

First, observe that the $\Z2$-SUSY transformation of $ S $ is given by
\begin{align} \label{deltaS}
\delta S &= \int dt\, \delta F_{011} \nn \\
&= \int dt \big[ (i\dot{F}_{011}-F_{110})\epsilon_{10}+(i\dot{F}_{010}+F_{101})\epsilon_{01} \big].
\end{align}
Next, we prove that if $ F_{k00}(t) = 0, $ then $ F_{\ell\alpha\beta}(t) = 0 $ for any $\alpha, \beta$ and $ \ell \geq k. $ 
One may have the identity from \eqref{CompFieldsTrans}
\[
 \delta F_{k00} = (-1)^{k+\Delta_{1}}F_{k10}\, \epsilon_{10}+(-1)^{k+\Delta_{2}} F_{k01}\,\epsilon_{01} = 0
\]
which means that $ F_{k10} = F_{k01} = 0. $ It follows that
\begin{align}
  \delta F_{k10} &= (-1)^{k+\Delta_{2}}\{ F_{k11}+ (k+1) F_{k+100}\} \epsilon_{01} =0,  \nonumber \\
\delta F_{k01} &= (-1)^{k+\Delta_{1}} \{ F_{k11}- (k+1) F_{k+100}\}\epsilon_{10} = 0.
\end{align}
Therefore, we get $ F_{k11} = F_{k+100} = 0. $ 
Repeating the same computation for $ F_{k+100} = 0, $ one may see $ F_{\ell\alpha\beta}(t) = 0. $
By this fact, the integrability condition $ F_{100}(t) = 0$ implies $ F_{110} = F_{101} = 0 $ so that the integrand of $ \delta S $ is a total derivative which shows $ \delta S = 0.$  
Namely, the integrability guarantees the $\Z2$-SUSY invariance.

We have shown the compatibility of $\Z2$-SUSY transformation and integrability of functions on $\Min$. The superfield under consideration carries an  infinite dimensional representation of the $\Z2$-SUSY algebra. 
Therefore, our next problem is to find constrains on superfields by which the superfield carries an irrep.

 %%%%%%%%%%%%%%%%%%%%%%%%%%%%%%%%%%%%%%%%%%%%%%%%%%%%%%%%%%%%%%%%%%%%%%
 \section{Irreps and integrability} \label{SEC:IrrepInteg}
  \subsection{Irreps  of minimal $\Z2$-SUSY algebra} \label{SEC:Irreps}
  \setcounter{equation}{0}

In this section, we establish the superfield presentation of the irreps of the $\Z2$-SUSY algebra \eqref{defz22spa} by imposing constraints on the superfield. 
We also discuss possible $\Z2$-SUSY invariant actions. 
  
Before discussing the constraints, it is necessary to clarify the irreps of $\Z2$-SUSY algebra   \eqref{defz22spa}. 
Here we present only the results and the detailed analysis is found in Appendix A. 
The $\Z2$-SUSY algebra  has two quadratic Casimir $ H^2 $ and $Z^2.$ 
They are constants for an irrep and the constants are denoted by $E^2, \lambda $ for $ H^2, Z^2,$ respectively. 
Therefore, each irrep is specified by two parameters $E^2$ and $\lambda. $ 
Irrep of the $\Z2$-SUSY algebra is four dimensional and there exist two types of them. 
One corresponds to $ \lambda  = 0 $ and $Z$ is represented trivially. 
The other corresponds to $ \lambda \neq 0 $ and $ Z$ is not trivial.

The physical meaning of $E$ is clear as $H$ is the Hamiltonian, while it is hard to see the physical meaning of $\lambda $ from \eqref{defz22spa}.  
In the known examples of $\Z2$-graded SUSY quantum mechanics,   
these parameters are constrained $ \lambda \neq 0 $ and $ \lambda = E^2 $ \cite{BruDup,AAD}. 
Thus, in the present work we restrict ourselves to the following two cases: 
(i) $\lambda = 0,$ (ii) $\lambda = E^2 \neq 0.$

The basis of the irreps is a set of differentiable functions of $t$ (time)
\begin{equation}
 \{ \underbrace{\phi(t)}_{(0,0)},\quad \underbrace{F(t)}_{(1,1)},\quad \underbrace{\psi(t)}_{(1,0)},\quad \underbrace{\xi(t)}_{(0,1)}\}.
\end{equation}
Because of this choice, the parameter $E$ becomes the time derivative $ E = i\frac{d}{dt} $ (see \S \ref{SEC:nuE}).  
The irreducible actions of the generators on this basis are given as follows:

\medskip\noindent
Case (i) $ \lambda = 0:$
   \begin{align}
     H 
     \begin{pmatrix}
        \phi \\ \; F \; \\ \psi \\ \xi
     \end{pmatrix}
     &= 
     E 
     \begin{pmatrix}
        \phi \\ \; F \; \\ \psi \\ \xi
     \end{pmatrix}
     =
     \begin{pmatrix}
        i\dot{\phi} \\ \; i\dot{F} \; \\ i\dot{\psi} \\ i\dot{\xi}
     \end{pmatrix},
     \nn \\
     Z
     \begin{pmatrix}
        \phi \\ \; F \; \\ \psi \\ \xi
     \end{pmatrix}
     &= 0,
     \nn \\
     Q_{10}
     \begin{pmatrix}
        \phi \\ \; F \; \\ \psi \\ \xi
     \end{pmatrix}
     &= 
      \begin{pmatrix}
        0 & 0 & 1 & 0\\
        0 & 0 & 0 & E \\
        E & 0 & 0 & 0 \\
        0 & 1 & 0 & 0
      \end{pmatrix}        
     \begin{pmatrix}
        \phi \\ \; F \; \\ \psi \\ \xi
     \end{pmatrix}
     =
     \begin{pmatrix}
       \psi \\ i\dot{\xi} \\ i\dot{\phi} \\ F
     \end{pmatrix},
     \nn \\
     Q_{01}
     \begin{pmatrix}
        \phi \\ \; F \; \\ \psi \\ \xi
     \end{pmatrix}
     &=
       \begin{pmatrix}
          0 & 0 & 0 & 1 \\
          0 & 0 & E & 0 \\
          0 & 1 & 0 & 0 \\
          E & 0 & 0 & 0
       \end{pmatrix}
     \begin{pmatrix}
        \phi \\ \; F \; \\ \psi \\ \xi
     \end{pmatrix}
     = 
     \begin{pmatrix}
      \xi \\ i\dot{\psi} \\ F \\ i \dot{\phi}
     \end{pmatrix}. \label{4D1paramatrix}
   \end{align}
Case (ii) $ \lambda = E^2 \neq 0:$
    \begin{align}
      H 
      \begin{pmatrix}
         \phi \\ \; F \; \\ \psi \\ \xi
      \end{pmatrix}
      &= 
      E 
      \begin{pmatrix}
         \phi \\ \; F \; \\ \psi \\ \xi
      \end{pmatrix}
      =
      \begin{pmatrix}
         i\dot{\phi} \\ \; i\dot{F} \; \\ i\dot{\psi} \\ i\dot{\xi}
      \end{pmatrix},
      \nn \\
      Z
      \begin{pmatrix}
         \phi \\ \; F \; \\ \psi \\ \xi
      \end{pmatrix}
      &= 
       \begin{pmatrix}
         0 & 1 & 0 & 0 \\
         E^2 & 0 & 0 & 0 \\
         0 & 0 & 0 & -iE \\
         0 & 0 & iE & 0   
       \end{pmatrix}
      \begin{pmatrix}
         \phi \\ \; F \; \\ \psi \\ \xi
      \end{pmatrix}
      =
      \begin{pmatrix*}[r]
         F \\ -\ddot{\phi} \\ \dot{\xi} \\ -\dot{\psi}
      \end{pmatrix*}
      \nn \\
      Q_{10}
      \begin{pmatrix}
         \phi \\ \; F \; \\ \psi \\ \xi
      \end{pmatrix}
      &= 
       \begin{pmatrix}
         0 & 0 & 1 & 0\\
         0 & 0 & 0 & iE \\
         E & 0 & 0 & 0 \\
         0 & -i & 0 & 0
       \end{pmatrix}        
      \begin{pmatrix}
         \phi \\ \; F \; \\ \psi \\ \xi
      \end{pmatrix}
      =
      \begin{pmatrix*}[r]
        \psi \\ -\dot{\xi} \\ i\dot{\phi} \\ -iF
      \end{pmatrix*},
      \nn \\
      Q_{01}
      \begin{pmatrix}
         \phi \\ \; F \; \\ \psi \\ \xi
      \end{pmatrix}
      &=
        \begin{pmatrix}
           0 & 0 & 0 & 1 \\
           0 & 0 & -iE & 0 \\
           0 & i & 0 & 0 \\
           E & 0 & 0 & 0
        \end{pmatrix}
      \begin{pmatrix}
         \phi \\ \; F \; \\ \psi \\ \xi
      \end{pmatrix}
      = 
      \begin{pmatrix*}[r]
       \xi \\ \dot{\psi} \\ iF \\ i \dot{\phi}
      \end{pmatrix*}.  \label{4DZneq0matrix}
    \end{align}

%%%%%%%%%%%%%%%%%%%%%%%%%%%%%%%%%%%%%%%%%%%%%%%%%%%%%%%%%%%%%%%%%%%%%%
 \subsection{Constraints on superfields ans $\Z2$-SUSY actions}
 
 We give the constraints on the superfield \eqref{CompExpansion} which reproduce the two types of irreps presented in \S \ref{SEC:Irreps}. 
%We consider the superfield of $\Z2$-degree $(0,0)$, the superfields of the other degree are treated in the same way. 
 
\medskip\noindent
Case (i): The constraint is given by  $ f_{100}(t) = 0.$ 

 This is nothing but the integrability condition for $ \Psi$ (see \S \ref{SEC:DerInt}). 
As shown in \S \ref{SEC:InvInt}, all the component fields vanish except the first four so that it is equivalent to the \textit{$z$-constraint} discussed by Bruce \cite{brusigma}. 
Thus the superfield becomes  the finite series
\begin{equation} \label{Case1Psi}
\Psi (t,z,\theta_{10},\theta_{01}) =  f_{000}(t)+\theta_{10}f_{010}(t)+ \theta_{01}f_{001}(t)+\theta_{10}\theta_{01}f_{011}(t)
\end{equation}
and it carries the four dimensional representation which is read off from \eqref{CompFieldsTrans}. 
For the $\Z2$-degree $(0,0)$ superfield, the SUSY transformation is given by 
\begin{align}
  \delta_{10} ( f_{000}, f_{011}, f_{010}, f_{001})
  &= (f_{010},  i\dot{f}_{001}, i \dot{f}_{000}, f_{011})\,\epsilon_{10},
  \nonumber \\
  \delta_{01} ( f_{000}, f_{011}, f_{010}, f_{001})
  &= ( f_{001}, i\dot{f}_{010}, f_{011}, i \dot{f}_{000})\,\epsilon_{01}. \label{z-const-action}
\end{align}
Making the identification $ ( f_{000}, f_{010}, f_{001}, f_{011}) = (\phi, \psi, \xi, F) $ and defining the action of $ Q_{10}, Q_{01}$ on the component fields by
\begin{equation} \label{Qrightaction}
  \delta \Psi = \sum_{k=0}^{\infty}\sum_{\alpha,\beta=0,1} z^{k}\,\theta_{10}^{\alpha}\,\theta_{01}^{\beta}\, \delta f_{k\alpha\beta} 
  =
  \Psi (\overleftarrow{Q}_{10} \epsilon_{10} + \overleftarrow{Q}_{01} \epsilon_{01}),
\end{equation}
it is immediate to see that the representation given by \eqref{z-const-action} is identical to the irrep \eqref{4D1paramatrix}.  
Thus, it has been shown that the superfield \eqref{Case1Psi} carries the $ \lambda = 0 $ irrep of  the $\Z2$-SUSY algebra.

Because the superfield $\Psi$ is independent of $z$, the analogue of the standard kinetic term $ D_{10}\Psi \cdot D_{01}\Psi $ and any function of $ \Psi $ are also integrable. 
Therefore, the action defined by
\begin{align}
  S &= -\int dt\, dz\, d\theta_{10}\, d\theta_{01} 
    \big( D_{10}\Phi_{00}\cdot D_{01}\Phi_{00} + \mu F(\Psi)\big) 
  \nonumber \\  
    &= \int dt \bigg( \dot{\phi}^{2}+ F^{2} +i\psi\dot{\psi}+i\xi \dot{\xi}  - \mu F_{011}(t)  \bigg)
\end{align}
is invariant under the $\Z2$-SUSY transformation. 
One need to introduce the coupling constant $\mu$ of $\Z2$-degree $(1,1)$ to make the integrand homogeneous. 
This is the same situation as the previous studies \cite{AKTcl,brusigma,bruSG}. 
The action consists of one propagating boson $\phi$, two propagating fermions $\psi, \xi$ 
and one auxiliary boson of degree $(1,1)$ (exotic boson).
This is because of the choice of the superfield of degree $(0,0).$ 
One may apply the constraint $ f_{100} = 0 $ to the superfields of non-trivial $ \Z2$-degree. 
Then the superfield carries an irrep of $\Z2$-SUSY algebra which is obtained from \eqref{4D1paramatrix} by the dressing transformation. 
Such superfield produces an invariant action with different structure. 
We present the detail in Appendix B.

\medskip\noindent
Case (ii): The constraint is given by  $ f_{011}(t) = 0.$

The $\Z2$-SUSY transformation of this constraint requires
\begin{equation}
\delta f_{011} = (-1)^{\Delta_{1}}\{ i\dot{f}_{001}- f_{110}\}\epsilon_{10}+(-1)^{\Delta_{2}}  \{ i\dot{f}_{010}+  f_{101}\}\epsilon_{01}=0.
\end{equation}
Thus we obtain
\begin{equation}
   i\dot{f}_{001}=f_{110}, \qquad i\dot{f}_{010}=-f_{101}.
\end{equation}
Applying the $\Z2$-SUSY transformation to the both sides of these relations, one has
\begin{align}
-i (-1)^{\Delta_{1}} \dot{f}_{100}\epsilon_{10}-(-1)^{\Delta_{2}} \ddot{f}_{000}\epsilon_{01}&= -(-1)^{\Delta_{1}}i \dot{f}_{100}\epsilon_{10}- (-1)^{\Delta_{2}}\{ f_{111}+ 2 f_{200}\}\epsilon_{01}, \nn \\
- (-1)^{\Delta_{1}}\ddot{f}_{000}\epsilon_{10}+i (-1)^{\Delta_{2}} \dot{f}_{100}\epsilon_{01} &= (-1)^{\Delta_{1}} \{ f_{111}- 2 f_{200}\}\epsilon_{10}+(-1)^{\Delta_{2}}i \dot{f}_{100}\epsilon_{01}.
\end{align}
These relations deduce the followings
\begin{equation}
   \ddot{f}_{000}=2f_{200}, \qquad f_{111}=0
\end{equation}
Repeating the same process, one may see easily that all the higher order component fields with respect to $z$ are time derivatives of  $f_{000},f_{010},f_{001},f_{100}$:
\begin{align}
f_{2k00}&=\frac{1}{(2k)!} f_{000}^{(2k)},  &
f_{2k+100}&=\frac{1}{(2k+1)!} f_{100}^{(2k)}, \nn \\
f_{2k10}&=\frac{1}{(2k)!} f_{010}^{(2k)},  &
f_{2k+110}&=\frac{i}{(2k+1)!} f_{001}^{(2k+1)}, \nn \\
f_{2k01}&=\frac{1}{(2k)!} f_{001}^{(2k)},  &
f_{2k+101}&=\frac{-i}{(2k+1)!} f_{010}^{(2k+1)}, \nn \\
f_{k11}&=0, &  k \geq 1&
\end{align}
where $f^{(n)}$ denotes the $n$-th time derivative of $f(t)$. 
Therefore, the $\Z2$-superfield is the finitely generated infinite series
\begin{align} \label{Case2Psi}
\Psi (t,z,\theta_{10},\theta_{01}) &=  f_{000}(t)+\theta_{10}f_{010}(t)+ \theta_{01}f_{001}(t)+zf_{100}(t) + z \theta_{10} i \dot{f}_{001} - z\theta_{01} i \dot{f}_{010}  \nn \\
  + \sum_{k=1}^{\infty} & \left[ \frac{z^{2k}}{(2k)!} f_{000}^{(2k)} 
   +\frac{z^{2k+1}}{(2k+1)!} f_{100}^{(2k)} 
   +\frac{z^{2k}\theta_{10}}{(2k)!} f_{010}^{(2k)}
   \right.
   \nn \\
   & \left.
    +\frac{iz^{2k+1}\theta_{10}}{(2k+1)!} f_{001}^{(2k+1)} 
   +\frac{z^{2k}\theta_{01}}{(2k)!} f_{001}^{(2k)}
    -\frac{iz^{2k+1}\theta_{01}}{(2k+1)!} f_{010}^{(2k+1)}  \right].
\end{align}
For the $\Z2$-degree $(0,0)$ superfield, the $\Z2$-SUSY transformation of the generating component fields is given by
\begin{align} \label{f011const-action}
  \delta_{10} (f_{000}, f_{100}, f_{010}, f_{001}) &= 
  (f_{010}, -i\dot{f}_{001}, i \dot{f}_{000}, -f_{100})\,\epsilon_{10},
  \nn \\
  \delta_{01} (f_{000}, f_{100}, f_{010}, f_{001}) &= 
  (f_{001}, i\dot{f}_{010}, f_{100}, i \dot{f}_{000})\,\epsilon_{01}. 
\end{align}
Making the identification 
$ (f_{000},f_{010},f_{001},f_{100}) = (\phi,\psi,\xi,iF) $ and noting \eqref{Qrightaction}, 
it is immediate to see that the representation given by \eqref{f011const-action} is identical to the irrep \eqref{4DZneq0matrix}.
Thus it has been shown that the superfield \eqref{Case2Psi} carries the $ \lambda \neq 0 $ irrep of  the $\Z2$-SUSY algebra.

The kinetic action invariant under the transformation \eqref{4DZneq0matrix} is given by 
\begin{equation}
   S = \int dt (-\dot{\phi}^2 + F^2 + i \psi \dot{\psi} + i \xi \dot{\xi}). 
\end{equation}
However, the standard kinetic term for the superfield \eqref{Case2Psi} is not integrable as   $ D_{10}\Psi \cdot D_{01} \Psi $ has the term $ -iz(\psi\dot{\psi}-\xi \dot{\xi}) $ and vanishing this term is not physically acceptable. 
Therefore, integration of $ D_{10}\Psi \cdot D_{01} \Psi $ on $\Min$ does not reproduce the invariant kinetic term.

 %%%%%%%%%%%%%%%%%%%%%%%%%%%%%%%%%%%%%%%%%%%%%%%%%%%%%%%%%%%%%%%%%%%%%%
\section{Concluding remarks}

For the minimal $\Z2$-SUSY, we investigated irreps of the algebra, superfield presentation, invariant actions and compatibility with integration on the $\Z2$-superspace $\Min.$ 
One of the irreps where $ Z$ represented non-trivially has the superfield presentation,  but it is not compatible with the integral. 
Of course, this does not rule out superfield presentation of the irrep since there are  possibilities of other constrains and the use of multiple or non-scalar type superfields.  
Rather, the lesson from this is that the superfield formalism of $\Z2$-SUSY requires more careful analysis than the standard SUSY. 

The present work deals with the minimal $\Z2$-SUSY which has only one supercharge in each sector of odd parity. 
Thus, it corresponds to $ {\cal N}=1$ standard SUSY. 
Most fundamental SUSY is $ {\cal N}=2 $ so that it is desirable to study ${\cal N}$-extended $\Z2$-SUSY in a way similar to the present work. 
An ${\cal N}$-extended SUSY quantum mechanics is discussed in \cite{AAD}. 
As is seen from the present work, the knowledge on irreps of ${\cal N}$-extended $ \Z2$-SUSY algebra is quite important but it is still an open problem. 
Neither superfield nor other formulation of ${\cal N}$-extended $\Z2$-SUSY classical theory has not studied in literature. These should be works to be done.

Differential geometry on $\mathbb{Z}_2^n$-supermanifold started in \cite{CGP1,CGP2} (for recent developments see \cite{BruIbarPon} and references therein) is one of the extensively studied topics in mathematics. 
It is well known that SUSY is a quite useful tool for differential geometry and differential topology (see e.g. \cite{Smilga}). 
Therefore, one may expect that development of $\Z2$-superfield formalism would contribute the development $\mathbb{Z}_2^n$-differential geometry from physics side.

%%%%%%%%%%%%%%%%%%%%%%%%%%%%%%%%%%%%%%%%%%%%%%%%%%%%%%%%%%%%%%%%%%%%%%
 \appendix
 \section{Irreps of $\Z2$-graded SUSY algebra}
 \setcounter{equation}{0}

In this appendix, we discuss irreps of the $\Z2$-SUSY algebra \eqref{defz22spa}. 
We denote this algebra simply by $\g.$ 
The degree $(0,0)$ element $H$ is the center (commute with all elements) of $\g $ and the degree $(1,1)$ element $Z$ is the $\Z2$-graded center ($\Z2$-commutative with all elements). 
It follows that $ H^2 $ and $ Z^2$ are the  quadratic Casimir elements which means that they commute all the elements of $\g.$  
The maximal Abelian subalgebra (Cartan subalgebra) of $\g $  is 
 $ \mathfrak{h} = \mathrm{lin.\ span}\; \langle \ H, \ Z \ \rangle. $ 
  
 We want irreps of $\g$ on a $\Z2$-graded vector space $V$
 \begin{equation}
    V = V_{(0,0)} \oplus V_{(1,1)} \oplus V_{(1,0)} \oplus V_{(0,1)}.
 \end{equation} 
Our strategy is to start with the irreps of the Cartan subalgebra $\mathfrak{h}$ and induce representations of $\g$ from them. 
The irrep of $ \mathfrak{h} $ is either one dimensional or two dimensional and they are given as follows \cite{AA}
   \begin{enumerate}
     \renewcommand{\labelenumi}{\normalfont (\roman{enumi})}
     \item One dimensional irrep on $ \nu(E) = \mathrm{lin.\ span}\; \langle \ \ket{00} \ \rangle $
       \begin{equation}
         H\ket{E} = E \ket{00}, \qquad Z \ket{00} = 0.
       \end{equation}
     \item Two dimensional irrep on $\nu(E,\lambda) = \mathrm{lin.\ span}\; \langle \ \ket{00}, \ \ket{11} \ \rangle $
       \begin{align}
          H\ket{00} &= E \ket{00}, \qquad \ket{11} = Z \ket{00}, \qquad
          Z \ket{11} = \lambda \ket{00}, \quad \lambda \neq 0
       \end{align}
    where $ E, \lambda $ are (in general complex) parameters and the $\Z2$-degree of $\ket{00} $ is supposed to be $(0,0).$ 
\end{enumerate}   
 %
 
 %%%%%%%%%%%%%%%%%%%%%%%%%%%%%%%%%%%%%%%%%%%%%%%%%%%%%%%%%%%%%%%%%%%%%%
 \subsection{Induced irrep from $\nu(E)$} \label{SEC:nuE}
 
 A four dimensional vector space $V(E)$ is induced from $\nu(E)$ whose basis is given by
 \begin{equation}
   \ket{00}, \quad   \ket{01}:= Q_{01} \ket{00}, \quad \ket{10} := Q_{10} \ket{00}, 
   \quad \ket{\widetilde{11}}:= \hf\{Q_{01}, Q_{10}\} \ket{00}
 \end{equation}
 where the argument of ket vector indicates its $\Z2$-degree and the tilde is to keep using the same notation as \S \ref{SEC:nuElam}. 
 Thus the each subspace with the fixed $\Z2$-degree is one dimensional. 
 
 The action of $\g$ on $V(E)$ is readily computed by the use of the defining relations \eqref{defz22spa}. 
 Define the four component vector 
 $ \Psi := (\; \ket{00}, \ket{\widetilde{11}}, \ket{10}, \ket{01} \;).$ 
 Then
 \begin{align} \label{1Dinduction}
    H \Psi &= E \Psi,
    \qquad
    Z \Psi = 0,
    \nn \\
    Q_{10} \Psi &= (\; \ket{10}, E\ket{01}, E\ket{00}, \ket{\widetilde{11}} \;),
    \nn \\
    Q_{01} \Psi &= (\; \ket{01}, E\ket{10},  \ket{\widetilde{11}}, E\ket{00} \;).
 \end{align}
 The eigenvalue of the Casimir element $H^2, Z^2$ is $E^2, 0, $ respectively. 
 Obviously, this is an irrep of $\g$. 
  
  In order to make a connection between the irrep on $V(E)$ and $\Z2$-graded classical mechanics, we make Fourier transform of the irrep basis:
 \begin{equation}
   \ket{00}_t := \frac{1}{\sqrt{2\pi}} \int dE e^{-iEt} \ket{00}, \quad \mathrm{etc.} 
   \label{FourierT}
 \end{equation}
 Then $H$ acts on these basis as the differential operator, e.g.
 \begin{equation}
   H \ket{00}_t  = i \frac{\partial}{\partial t}  \ket{00}_t. 
 \end{equation} 
 We change the notation of the basis from ket  to a function of $t$
 \begin{equation}
   \begin{pmatrix}
      \ket{00}_t \\
      \; \ket{\widetilde{11}}_t \; \\
      \ket{10}_t \\
      \ket{01}_t
   \end{pmatrix}
   \ \to \ 
   \begin{pmatrix}
      \phi(t) \\ \; F(t) \; \\ \psi(t) \\ \xi(t)
   \end{pmatrix}.
 \end{equation}
 Then the action \eqref{1Dinduction} gives the irrep \eqref{4D1paramatrix}. 
 
 %%%%%%%%%%%%%%%%%%%%%%%%%%%%%%%%%%%%%%%%%%%%%%%%%%%%%%%%%%%%%%%%%%%%%%
 \subsection{Induced irrep from $\nu(E,\lambda)$} \label{SEC:nuElam}
 
 The following vectors are constructed on $ \nu(E,\lambda):$
 \begin{alignat}{2}
   \ket{00} &, & \qquad\qquad \ket{\widetilde{00}} &:= \hf\{Q_{01}, Q_{10} \} \ket{11},
   \nn \\
   \ket{11} &= Z \ket{00}, & \ket{\widetilde{11}} &= \hf\{Q_{01}, Q_{10} \} \ket{00},
   \nn \\
   \ket{10} &:= Q_{10} \ket{00}, & \ket{\widetilde{10}} &:= Q_{01} \ket{11},
   \nn \\
   \ket{01} &:= Q_{01} \ket{00}, & \ket{\widetilde{01}} &:= Q_{10} \ket{11}. 
   \label{basisVEl}
 \end{alignat}
 We suppose that these vectors form a basis of an eight dimensional vector space which is denoted by $V(E,\lambda).$ 
All the vectors are the simultaneous eigenvectors of $ H, H^2, Z^2 $ with the eigenvalues $ E, E^2, \lambda,$ respectively. 
 One may also compute the action of $Q_{10}, Q_{01} $ and $Z$ on these vectors. 
 They are given by
 \begin{alignat}{2}
   Q_{10} \ket{00} &= \ket{10}, & \qquad
   Q_{10} \ket{\widetilde{00}} &=  i \lambda \ket{10} + E \ket{\widetilde{10}},
   \nn \\
   Q_{10} \ket{11} &= \ket{\widetilde{01}}, & 
   Q_{10} \ket{\widetilde{11}} &= E\ket{01} + i \ket{\widetilde{01}},
   \nn \\
   Q_{10} \ket{10} &= E \ket{00}, & 
   Q_{10} \ket{\widetilde{10}} &= -i \lambda \ket{00} + \ket{\widetilde{00}},
   \nn \\
   Q_{10} \ket{01} &= -i\ket{11} + \ket{\widetilde{11}}, & 
   Q_{10} \ket{\widetilde{01}} &= E \ket{11}
 \end{alignat}
 and
 \begin{alignat}{2}
   Q_{01} \ket{00} &= \ket{01}, & \qquad
   Q_{01} \ket{\widetilde{00}} &= -i\lambda \ket{01} + E \ket{\widetilde{01}},
   \nn \\
   Q_{01} \ket{11} &= \ket{\widetilde{10}}, & 
   Q_{01} \ket{\widetilde{11}} &= E\ket{10} - i \ket{\widetilde{10}},
   \nn \\
   Q_{01} \ket{10} &= i \ket{11} + \ket{\widetilde{11}}, & 
   Q_{01} \ket{\widetilde{10}} &= E \ket{11},
   \nn \\
   Q_{01} \ket{01} &= E\ket{00}, & 
   Q_{01} \ket{\widetilde{01}} &= i\lambda \ket{00} + \ket{\widetilde{00}}
 \end{alignat}
 and 
 \begin{alignat}{2}
   Z \ket{00} &= \ket{11}, & \qquad
   Z \ket{\widetilde{00}} &= \lambda \ket{\widetilde{11}},
   \nn \\
   Z \ket{11} &= \lambda \ket{00},  & 
   Z \ket{\widetilde{11}} &= \ket{\widetilde{00}},
   \nn \\
   Z \ket{10} &= -\ket{\widetilde{01}}, & 
   Z \ket{\widetilde{10}} &= -\lambda \ket{01},
   \nn \\
   Z \ket{01} &= -\ket{\widetilde{10}}, & 
   Z \ket{\widetilde{01}} &= -\lambda \ket{10}.
 \end{alignat}
 
 The vector space $V(E,\lambda)$ carries a reducible representation of $\g $ since one may show the existence of the invariant subspace $W(E,\lambda) \subset V(E,\lambda).$ 
The space $ W(E,\lambda)$ is also an $\Z2$-graded vector space:
 \begin{equation}
   W = W_{(0,0)} \oplus W_{(1,1)} \oplus W_{(1,0)} \oplus W_{(0,1)}.
 \end{equation}
 Note that if one of the subspaces, say $ W_{(0,0)},$ is of two dimension, then $ W(E,\lambda) = V(E, \lambda) $ since other subspaces are also two dimensional as they are obtained by the action of $\g.$ 
 Therefore,  $ \dim W_{(a,b)} = 1$ for all $ (a,b).$  
 Now suppose that
 \begin{equation}
   \ket{v_{00}} = \alpha \ket{00} + \beta \ket{\widetilde{00}} \in W_{(0,0)}, 
   \quad \alpha, \beta \in \mathbb{C}
 \end{equation}
 Then, two vectors
 \begin{align}
    Q_{10} \ket{v_{00}} &= (\alpha+i\lambda \beta) \ket{10} + \beta E \ket{\widetilde{10}},
    \nn \\
    Q_{01} Z\ket{v_{00}} &= \lambda \beta E \ket{10} + (\alpha-i\lambda \beta) \ket{\widetilde{10}}   
 \end{align}
 belong to $ W_{(1,0)}$ so that there must be a constant $ c \in \mathbb{C}$ such that
 \begin{equation}
    Q_{01} Z \ket{v_{00}} = c\, Q_{10} \ket{v_{00}}.  \label{Q01ZQ10}
 \end{equation}
 From \eqref{Q01ZQ10} one may readily obtain
 \begin{align}
    Q_{10} Q_{01} Z \ket{v_{00}} &= c E \ket{v_{00}} \in W_{(0,0)},
    \nn\\
    Q_{10}Z \ket{v_{00}} &= \frac{\lambda}{c} Q_{01} \ket{v_{00}} \in W_{(0,1)},
    \nn \\
    Q_{01} Q_{10} \ket{v_{00}} &= \frac{E}{c}Z \ket{v_{00}} \in W_{(1,1)}.
 \end{align}
 The relation \eqref{Q01ZQ10} is equivalent to
 \begin{equation}
    \lambda \beta E = c (\alpha +i \lambda \beta),
    \qquad
    \alpha - i \lambda \beta = c E \beta
 \end{equation}
 which gives
 \begin{align}
    \alpha^2 &= \lambda \beta^2 (E^2-\lambda),
    \nn \\
    (Ec + i\lambda)^2 &= \lambda (E^2-\lambda).  \label{alpha2}
 \end{align}
 Therefore, for a given pair  $ (E, \lambda),$ there always exist  $ \alpha, \beta $ and $c$ which show the existence of four dimensional invariant subspace 
 \[ W(E,\lambda) = \mathrm{lin.\ span}\; \langle \  \ket{v_{00}},\; \ket{v_{11}}:= Z \ket{v_{00}},\; \ket{v_{10}}:= Q_{10} \ket{v_{00}},\; \ket{v_{01}}:= Q_{01} \ket{v_{00}} \ \rangle. 
 \]
 $ W(E,\lambda)$ carries a four dimensional irrep of $ \g $ with two parameters $ E, \lambda. $  
 The action of $\g$ on $W(E,\lambda)$ is  summarized as follows:
 \begin{alignat}{2}
   H \ket{v_{ab}} &= E \ket{v_{ab}},
   \nn \\
   Q_{10} \ket{v_{00}} &= \ket{v_{10}}, & \qquad
   Q_{10} \ket{v_{11}} &= \frac{\lambda}{c} \ket{v_{01}},
   \nn \\
   Q_{10} \ket{v_{10}} &= E \ket{v_{00}}, & 
   Q_{10} \ket{v_{01}} &= \frac{cE}{\lambda} \ket{v_{11}},
   \nn \\
   Q_{01} \ket{v_{00}} &= \ket{v_{01}}, & 
   Q_{01} \ket{v_{11}} &= c \ket{v_{10}},
   \nn \\
   Q_{01} \ket{v_{10}} &= \frac{E}{c} \ket{v_{11}}, & 
   Q_{01} \ket{v_{01}} &= E \ket{v_{00}},
   \nn \\
   Z \ket{v_{00}} &= \ket{v_{11}}, & 
   Z \ket{v_{11}} &= \lambda \ket{v_{00}}, 
   \nn \\
   Z \ket{v_{10}} &= -\frac{\lambda}{c} \ket{v_{01}}, & 
   Z \ket{v_{01}} &= -c \ket{v_{10}}. \label{4D2paramirrep}
 \end{alignat}

 We now discuss the case in which the vectors \eqref{basisVEl} are not linearly independent. 
 This implies $ \ket{v_{00}} = 0$ for some non-vanishing $\alpha, \beta,$ i.e., one may have  
 \begin{equation}
   \ket{\widetilde{00}} = \mu \ket{00}.   \label{tilde00200}
 \end{equation}
 By the action of $ Q_{10}, Q_{01}, Z$ one has the relations:
 \begin{equation}
   \ket{\widetilde{10}} = \frac{1}{E}(-i\lambda + \mu) \ket{10},
   \quad
   \ket{\widetilde{01}} = \frac{1}{E}(i\lambda + \mu) \ket{01},
   \quad
   \ket{\widetilde{11}}=\frac{\mu}{\lambda} \ket{11}. \label{8D24D}
 \end{equation}
 Thus we end up with the four dimensional vector space $ \widetilde{V}(E,\lambda)$ spanned by
 \begin{equation}
   \ket{00}, \qquad \ket{11}, \qquad \ket{10}, \qquad \ket{01}. 
 \end{equation}
 Computing the action of $ Q_{10}Q_{01}Z$ on \eqref{tilde00200} we have
 \begin{equation}
      \mu^2 = \lambda (E^2-\lambda). 
 \end{equation}
 Recalling the relation \eqref{alpha2}, $\mu $ and $c$ of \eqref{Q01ZQ10} are related:
 \begin{equation}
    Ec =  \mu - i\lambda. 
 \end{equation}
 With this relation, one may verify that the action of $\g$ on $ \widetilde{V}(E,\lambda)$ gives the same irrep as on  $ W(E,\lambda).$ 
 Namely, \eqref{4D2paramirrep} with the replacement $ \ket{v_{ab}} \to \ket{ab}.$ 
 
  The existence of this four dimensional irrep with two parameters has not been pointed out in the literature, so that one of the novel results of the present work. 
For the particular choice of the parameters $ \lambda= E^2, $   we have $ \mu =0 $ and $ c = -i E.$ 
 Thus \eqref{4D2paramirrep} yields
 \begin{alignat}{2}
   H \ket{v_{ab}} &= E \ket{v_{ab}},
   \nn \\
   Q_{10} \ket{v_{00}} &= \ket{v_{10}}, & \qquad
   Q_{10} \ket{v_{11}} &= iE \ket{v_{01}},
   \nn \\
   Q_{10} \ket{v_{10}} &= E \ket{v_{00}}, & 
   Q_{10} \ket{v_{01}} &= -i \ket{v_{11}},
   \nn \\
   Q_{01} \ket{v_{00}} &= \ket{v_{01}}, & 
   Q_{01} \ket{v_{11}} &= -iE \ket{v_{10}},
   \nn \\
   Q_{01} \ket{v_{10}} &= i \ket{v_{11}}, & 
   Q_{01} \ket{v_{01}} &= E \ket{v_{00}},
   \nn \\
   Z \ket{v_{00}} &= \ket{v_{11}}, & 
   Z \ket{v_{11}} &= E^2 \ket{v_{00}}, 
   \nn \\
   Z \ket{v_{10}} &= -iE \ket{v_{01}}, & 
   Z \ket{v_{01}} &=iE \ket{v_{10}}. \label{4D2paramirrep2}
 \end{alignat} 
 We also present \eqref{8D24D} for this choice of $\lambda$ 
 \begin{equation}
   \ket{\widetilde{00}} = \ket{\widetilde{11}} = 0, 
   \qquad
   \ket{\widetilde{10}} = -iE \ket{10},
   \qquad
   \ket{\widetilde{01}} = iE \ket{01}.
 \end{equation}

 To have the classical fields, the Fourier transform of the irrep basis is made in a same way as \eqref{FourierT}. 
 The resulted $\Z2$-graded supermultiplet are denoted by the same notation as \S \ref{SEC:nuE} and it is straightforward to see that the $\Z2$-SUSY transformation is given by \eqref{4DZneq0matrix}.   
This $\Z2$-graded supermultiplet would corresponds to $(1,2,1)_{[00]}$ supermultiplet introduced in \cite{AKTcl}.  
 
 %%%%%%%%%%%%%%%%%%%%%%%%%%%%%%%%%%%%%%%%%%%%%%%%%%%%%%%%%%%%%%%%%%%%%%%%%%%%%%%
\section{Non-trivial $\Z2$-degree superfields presentation of irreps and invariant actions}
\setcounter{equation}{0}

In this appendix, we consider the superfields with non-trivial $\Z2$-degree under the constraint $ f_{100}(t) = 0.$ 
They also carry Case (i) $ \lambda = 0$ irreps of the $\Z2$-SUSY algebra  obtained from  \eqref{4D1paramatrix} by the dressing transformation discussed in \cite{PT,KRT}. 
For any superfield, we keep using the notation $ \phi(t), F(t), \psi(t), \xi(t) $ to indicate the degree $ (0,0), (1,1), (1,0), (0,1) $ component fields, respectively. 

\subsection{$(1,1)$ superfield}

The degree $(1,1)$ superfield is given by 
\begin{align}
\Psi_{11}(t,z,\theta_{10},\theta_{01}) = F(t)+\theta_{10}\xi(t)+\theta_{01}\psi(t)+\theta_{10}\theta_{01}\phi(t).
\end{align}
We see from \eqref{CompFieldsTrans} that the component fields transform under the $\Z2$-SUSY transformation as
\begin{align}
  Q_{10}
    \begin{pmatrix}
     \phi \\ \; F \; \\ \psi \\ \xi
  \end{pmatrix}
  &= 
   \begin{pmatrix}
0 & 0 & -E & 0\\
      0 & 0 & 0 & -1 \\
      -1 & 0 & 0 & 0 \\
      0 & -E & 0 & 0
   \end{pmatrix}        
  \begin{pmatrix}
     \phi \\ \; F \; \\ \psi \\ \xi
  \end{pmatrix}
  = -
  \begin{pmatrix*}
     i\dot{\psi} \\ \xi \\ \phi \\ i\dot{F} 
  \end{pmatrix*},
  \nn \\
  Q_{01}
    \begin{pmatrix}
     \phi \\ \; F \; \\ \psi \\ \xi
  \end{pmatrix}
  &=
    \begin{pmatrix}
0 & 0 & 0 & -E \\
        0 & 0 & -1 & 0 \\
        0 & -E & 0 & 0 \\
        -1 & 0 & 0 & 0
    \end{pmatrix}
  \begin{pmatrix}
 \phi \\ \; F \; \\ \psi \\ \xi
  \end{pmatrix}
  = -
  \begin{pmatrix*}
   i\dot{\xi} \\ \psi \\ i\dot{F} \\ \phi
  \end{pmatrix*}.  \label{11SF1para}
\end{align}
$ H = i \partial_t $ and $ Z = 0 $ as it should be. 
The representation matrices of $Q_{10}, Q_{01}$ given in \eqref{11SF1para} are obtained from those in \eqref{4D1paramatrix} by the dressing transformation
\begin{align}
   Q_{\vec{a}} &= U_1  Q_{\vec{a}}^{(0,0)} U_1^{-1}, \quad \vec{a} = (1,0),\; (0,1)
   \nn \\
   U_1 &= \mathrm{diag}(E^2,1,-E,-E).
\end{align}
This transformation is well-defined despite the presence of $ U_1^{-1}$ since the transformed matrix remains differential operator. 

The kinetic action invariant under \eqref{11SF1para} is given by
\begin{align}
 S&= \int  dt\, dz\, d\theta_{10}\, d\theta_{01}\, (D_{10}\Psi_{11})(D_{01}\Psi_{11}) \nn \\
 &= \int dt (\phi^{2}+\dot{F}^{2}+i\psi\dot{\psi}+i\xi\dot{\xi})
\end{align}
The physical boson is auxiliary and the exotic boson is propagating in this model. 
One may easily add interaction terms to the action, for instance
\begin{equation}
   \mu_1 F(\Psi_{11}) + \mu_2 G(\Psi_{11})
\end{equation}
where $ F(\Psi_{11}), G(\Psi_{11}) $  are odd and even functions of $ \Psi_{11}, $ respectively. 
The coupling constant $ \mu_1 $ and $ \mu_2$ has degree $ (0,0)$ and $(1,1)$, respectively.

 \subsection{$(1,0)$ superfield}
The degree $(1,0)$ superfield is given by   
\begin{align}
  \Psi_{10}(t,z,\theta_{10},\theta_{01}) = \psi(t)+\theta_{10}\phi(t)+\theta_{01}F(t)+\theta_{10}\theta_{01}\xi(t).
  \end{align}
Its component fields transform as  
\begin{align}
  Q_{10}
  \begin{pmatrix}
     \phi \\ \; F \; \\ \psi \\ \xi
  \end{pmatrix}
  &= 
   \begin{pmatrix}
         0 & 0 & -E & 0\\
         0 & 0 & 0 & -1 \\
         -1 & 0 & 0 & 0 \\
         0 & -E & 0 & 0
   \end{pmatrix}        
  \begin{pmatrix}
     \phi \\ \; F \; \\ \psi \\ \xi
  \end{pmatrix}
  = -
  \begin{pmatrix*}
      i\dot{\psi} \\ \xi \\ \phi \\ i\dot{F}
  \end{pmatrix*},
  \nn \\
  Q_{01}
  \begin{pmatrix}
     \phi \\ \; F \; \\ \psi \\ \xi
  \end{pmatrix}
  &=
    \begin{pmatrix}
0 & 0 & 0 & 1 \\
           0 & 0 & E & 0 \\
           0 & 1 & 0 & 0 \\
           E & 0 & 0 & 0
    \end{pmatrix}
  \begin{pmatrix}
 \phi \\ \; F \; \\ \psi \\ \xi
  \end{pmatrix}
  = 
  \begin{pmatrix*}
  \xi \\ i\dot{\psi} \\ F \\ i\dot{\phi}
  \end{pmatrix*}  \label{10SF1para}
\end{align}
and $ H = i\partial_t, Z = 0. $ 
The representation matrices of $Q_{10}, Q_{01}$ given in \eqref{10SF1para} are obtained from those in \eqref{4D1paramatrix} by the dressing transformation
\begin{align}
   Q_{\vec{a}} &= U_2  Q_{\vec{a}}^{(0,0)} U_2^{-1}, \quad \vec{a} = (1,0),\; (0,1)
   \nn \\
   U_2 &= \mathrm{diag}(E,-1,-1,E).
\end{align}
The kinetic action invariant under \eqref{10SF1para} is given by
\begin{align}
 S&=\int  dt\,dz\, d\theta_{10}\, d\theta_{01} (D_{10}D_{01}\Psi_{10})(H\Psi_{10}) \nn \\
 &= \int dt (\dot{\phi}^{2}-\dot{F}^{2}+i\dot{\psi}\ddot{\psi}-i\xi\dot{\xi})
 \end{align}
Both physical and exotic bosons are propagating in this model. 
However, it has higher order derivative of the degree $(1,0)$ fermion.

\subsection{$(0,1)$ superfield}

The degree $(0,1)$ superfield is given by       
\begin{align}
     \Psi_{01}(t,z,\theta_{10},\theta_{01}) =\xi(t) +\theta_{10}F(t)+\theta_{01}\phi(t)+\theta_{10}\theta_{01}\psi(t).
\end{align}
Its component fields transform as  
\begin{align}
  Q_{10}  
  \begin{pmatrix}
     \phi \\ \; F \; \\ \psi \\ \xi
  \end{pmatrix}
  &= 
   \begin{pmatrix}
               0 & 0 & 1 & 0\\
               0 & 0 & 0 & E \\
               E & 0 & 0 & 0 \\
               0 & 1 & 0 & 0
   \end{pmatrix}        
  \begin{pmatrix}
    \phi \\ \; F \; \\ \psi \\ \xi
  \end{pmatrix}
  =
  \begin{pmatrix*}
      \psi \\ i\dot{\xi} \\ i\dot{\phi} \\ F
  \end{pmatrix*},
  \nn \\
  Q_{01}
    \begin{pmatrix}
     \phi \\ \; F \; \\ \psi \\ \xi
  \end{pmatrix}
  &=
    \begin{pmatrix}
                 0 & 0 & 0 & -E \\
                 0 & 0 & -1 & 0 \\
                 0 & -E & 0 & 0 \\
                 -1 & 0 & 0 & 0
    \end{pmatrix}
  \begin{pmatrix}
 \phi \\ \; F \; \\ \psi \\ \xi
  \end{pmatrix}
  = -
  \begin{pmatrix*}
  i\dot{\xi} \\ \psi \\ i\dot{F} \\ \phi
  \end{pmatrix*}.  \label{01SF1para}
\end{align}
and $ H = i\partial_t, Z = 0. $ 
The representation matrices of $Q_{10}, Q_{01}$ given in \eqref{01SF1para} are obtained from those in \eqref{4D1paramatrix} by the dressing transformation
\begin{align}
   Q_{\vec{a}} &= U_3  Q_{\vec{a}}^{(0,0)} U_3^{-1}, \quad \vec{a} = (1,0),\; (0,1)
   \nn \\
   U_2 &= \mathrm{diag}(E,-1,E,-1).
\end{align}
The kinetic action invariant under \eqref{01SF1para} is given by
 \begin{align}
 S&= \int dt d\theta_{10} d\theta_{01} (D_{10}D_{01}\Psi_{10})(H\Psi_{10}) \nn \\
 &= \int dt (\dot{\phi}^{2}-\dot{F}^{2}-i\psi\dot{\psi}+i\dot{\xi}\ddot{\xi})
 \end{align}
Contrary to $(1,0)$ superfield, this model has higher derivative of the degree $(0,1)$ fermion.  
 
%%%%%%%%%%%%%%%%%%%%%%%%%%%%%%%%%%%%%%%%%%%%%%%%%%%%%%%%%%%%%%%%%%%%%%%%%%%%%%%
%
%  \section{References}
%
%%%%%%%%%%%%%%%%%%%%%%%%%%%%%%%%%%%%%%%%%%%%%%%%%%%%%%%%%%%%%%%%%%%%%%%%%%%%%
%

\end{document}